\documentclass[twocolumn,showpacs,preprintnumbers,amsmath,amssymb]{revtex4}
\usepackage{graphicx}% Include figure files
\usepackage{dcolumn}% Align table columns on decimal point
\usepackage{bm}% bold math

\begin{document}

%\preprint{APS/123-QED}

\title{Phase diagram of epidemic spreading - unimodal vs. bimodal probability distributions}% Force line breaks with \\

\author{Alen Lan\v{c}i\'{c}$^{a}$}%
\email{alen@student.math.hr}
\author{Nino Antulov-Fantulin$^{b}$}
\email{nino.antulov-fantulin@fer.hr}
\author{Mile \v{S}iki\'{c}$^{b}$}
\email{mile.sikic@fer.hr}
\affiliation{$^{a}$Faculty of Science, Department of Mathematics, \\
$^{b}$Faculty of Electrical Engineering and Computing, Department of Electronic Systems and Information Processing, \\
University of Zagreb, Zagreb, Croatia}
\author{Hrvoje \v{S}tefan\v{c}i\'{c}}
\email{shrvoje@thphys.irb.hr}
\affiliation{Theoretical Physics Division, Rudjer Bo\v{s}kovi\'{c} Institute, P.O.B. 180, HR-10002 Zagreb, Croatia}

\date{\today}% It is always \today, today,

\begin{abstract}
The disease spreading on complex networks is studied in SIR model. Simulations on empirical complex networks reveal two specific regimes of disease spreading: local containment and epidemic outbreak. The variables measuring the extent of disease spreading are in general characterized by a bimodal probability distribution. Phase diagrams of disease spreading for empirical complex networks are introduced. A theoretical model of disease spreading on m-ary tree is investigated both analytically and in simulations. It is shown that the model reproduces qualitative features of phase diagrams of disease spreading observed in empirical complex networks. The role of tree-like structure of complex networks in disease spreading is discussed. 
\end{abstract}

\maketitle

\section{Introduction}

The spreading of an epidemic on complex networks has been a subject of intensive research during the last decade \cite{rev1,rev2,rev3}. The importance of this line of research is evident both in disease control and prevention and the spreading of all forms of malicious software in computer and communication networks. A crucial element of a successful epidemic model is a good description of epidemic spreading pathways. Complex networks were found to be a good description of social contact networks \cite{collab,soc}, whereas the physical structure of IT and communication networks directly qualifies them as complex network systems \cite{internet1,internet2}. The ``paradigmatic'' characteristics of complex networks such as ``small world network'' property \cite{smallworld} and ``scale-free network'' property \cite{scalefree1,scalefree2} profoundly influence the patterns of the epidemic spreading. The roles of intercontinental air travel and the existence of highly connected hubs in the onset and fast spreading of a pandemic are distinctly important \cite{dd1,dd2}.

Epidemiological models, such as the SIR model used in this paper \cite{sir}, describe the stochastic process of disease spreading along the complex network pathways. The model parameters are $p$, the probability per time step of a node to get infected if its neighboring node is also infected and $q$, the probability per time step of an infected node to recover. A prominent question of the epidemiological model dynamics is the existence of thresholds for the onset of the epidemic.  In a model of homogeneous mixing based on the mass action principle a condition for the onset of the epidemic is formulated in terms of the {\it basic reproduction ratio $R_0$}, as $R_0>1$. On the other hand, studies of the disease spreading on scale-free complex networks in SIS model showed the absence of the epidemic threshold \cite{Vespignani}. In this paper, a reasonable notion that the structure of the disease spreading pathways influences the conditions for the onset of epidemic is elaborated and quantified.  
From the very stochastic nature of the SIR model it is easy to single out two specific patterns/regimes of the disease spreading: the disease may ``die-out'' after a (small) number of steps or the epidemic may spread throughout the entire network. However, in empirical complex networks the transition between these two regimes is not abrupt. There exists a considerable segment of parametric space in which the said regimes coexist, i.e. there is a non-negligible probability for the appearance of both of them. In this paper we study the conditions for the appearance of these regimes, examine the possibility of their coexistence and construct the phase diagram of epidemic spreading. Furthermore, we propose a simple theoretical model which qualitatively reproduces the observed phase diagrams. 

\section{Phase diagram of epidemic spreading}

The simulations of disease spreading in the SIR model on empirical networks reveal an interesting interplay between the network structure and the SIR model parameters. The variables used to measure the extent of disease spreading are the number of infected nodes and the epidemic range. The number of infected nodes is defined as the number of nodes that got infected at any moment during the spreading of the disease. The epidemic range is defined as a maximal number of steps that the infection travelled from the initially infected node. We study the disease spreading on a  number of empirical complex networks and in particular we present detailed results of simulations for the complex network of 2003 Condensed matter collaborations introduced in \cite{collab}.

\begin{figure}[t]
\centering
\includegraphics*[width=0.5\textwidth]{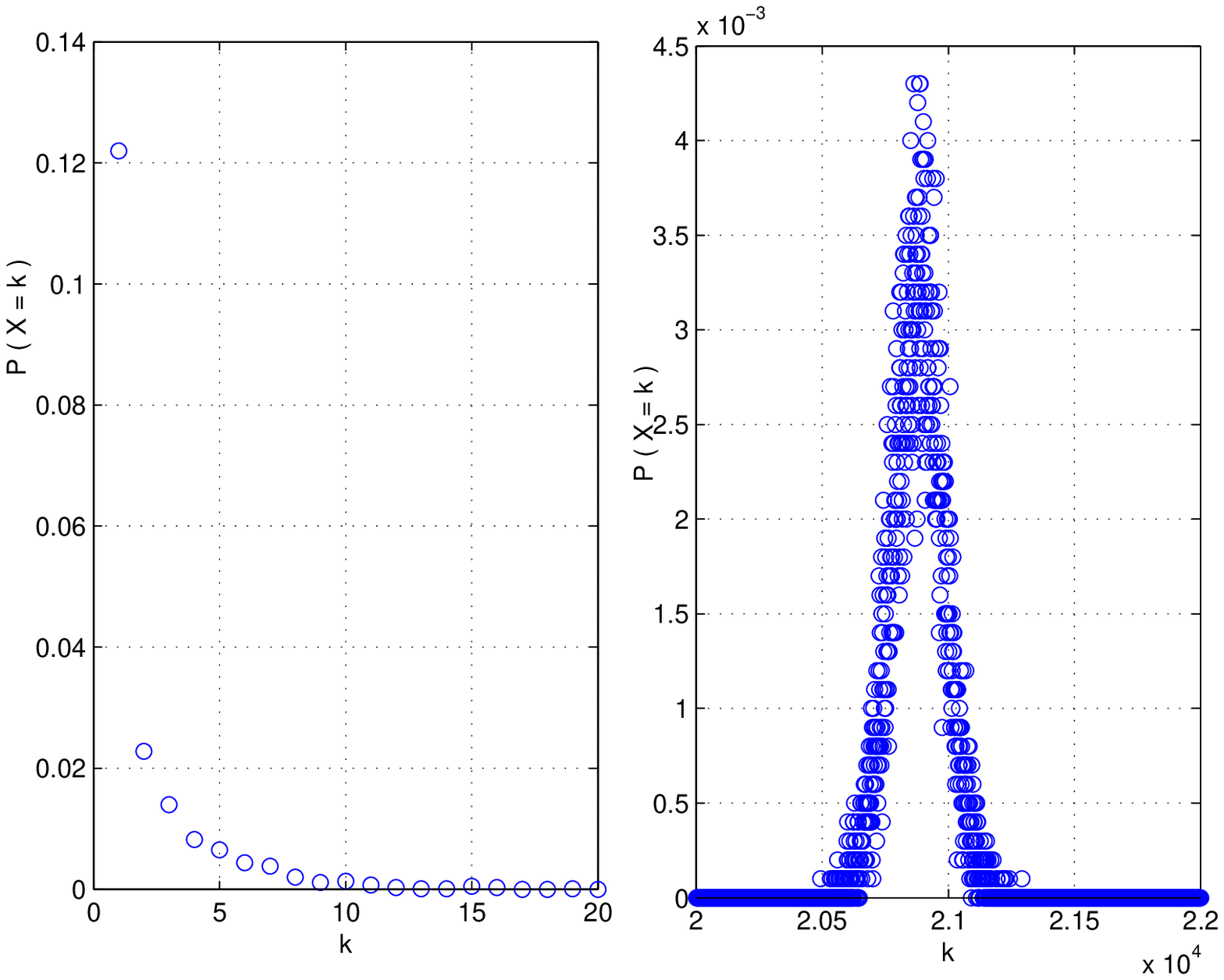}% Here is how to import EPS art
\caption{\label{fig:1} The number of infected nodes for network \cite{collab} with $p=0.3$ and $q=0.7$ averaged over 10000 simulations.} 
\end{figure}

\begin{figure}[t]
\centering
\includegraphics*[width=0.5\textwidth]{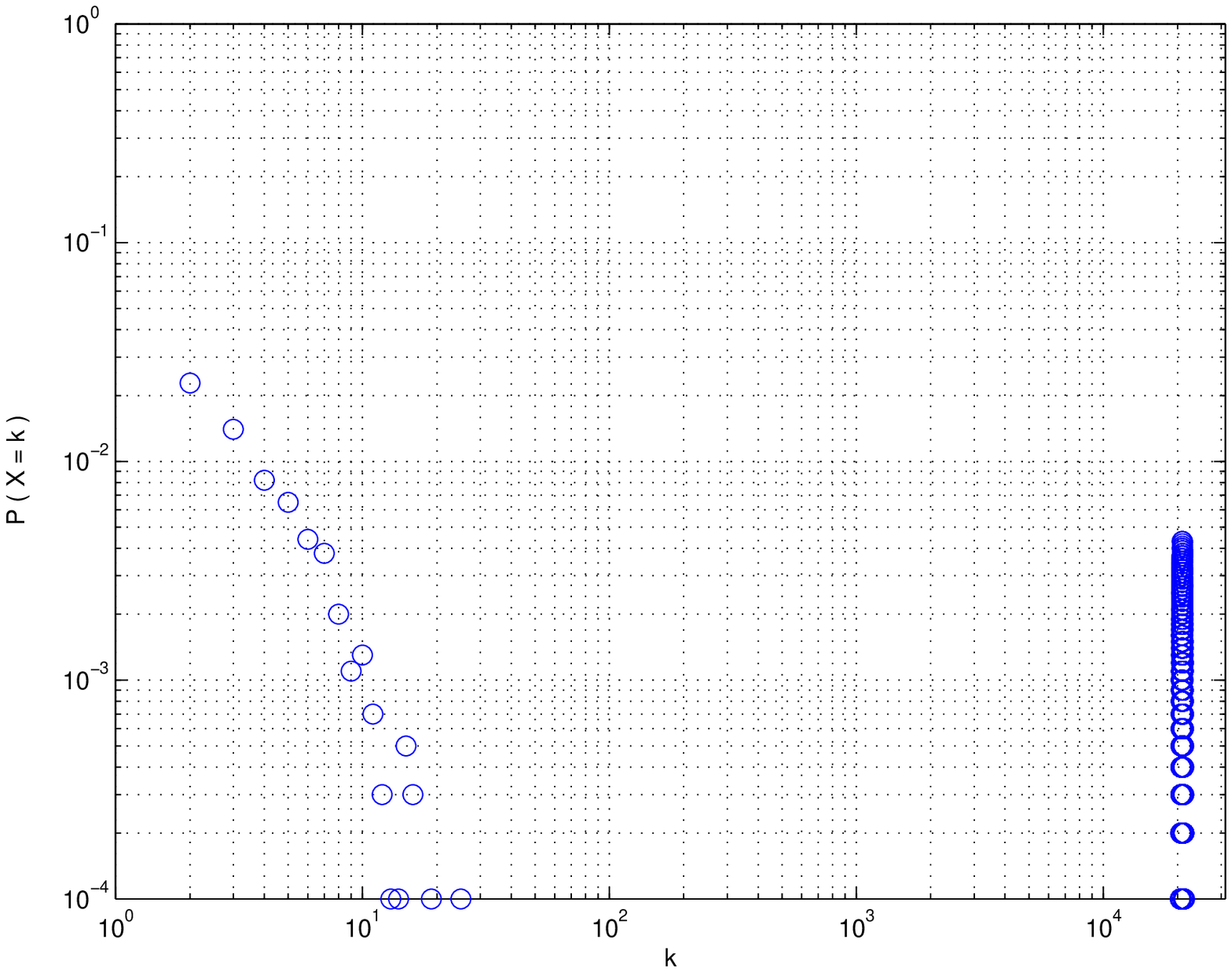}% Here is how to import EPS art
\caption{\label{fig:2} The number of infected nodes for network \cite{collab} with $p=0.3$ and $q=0.7$ averaged over 10000 simulations.}
\end{figure}

\begin{figure}[t]
\centering
\includegraphics*[width=0.5\textwidth]{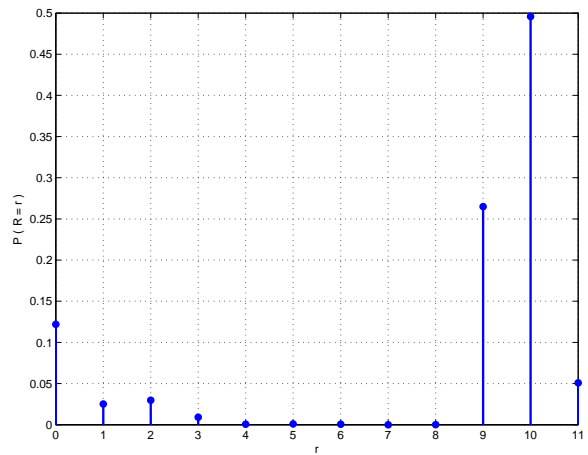}% Here is how to import EPS art
\caption{\label{fig:3} The distribution of the epidemic range for network \cite{collab} with $p=0.3$ and $q=0.7$ averaged over 10000 simulations. The maximal distance from the initially infected node is 11.}
\end{figure}

Let us start with an interesting observed feature which is central for the purpose of this paper: the probability distribution for the number of infected nodes is bimodal for some values of model parameters $p$ and $q$, as presented in Fig. \ref{fig:1} and Fig. \ref{fig:2}. This interesting feature was already studied and reported for the class of simulated scale-free and small world networks in \cite{gallos}. We further observe the bimodal probability distribution for the epidemic range as shown in Fig \ref{fig:3}. This coincidence in character of probability distributions strongly suggests that the bimodality is an important feature of the disease spreading in SIR model on the said complex network. Our goal in this paper is to find out to which extent the observed behavior is generic, i.e. at least qualitatively the same for different SIR model parameters, initially infected nodes and even for different choices of complex networks.

%{\bf stress that everything is practically bimodal}

\begin{figure}[t]
\centering
\includegraphics*[width=0.5\textwidth]{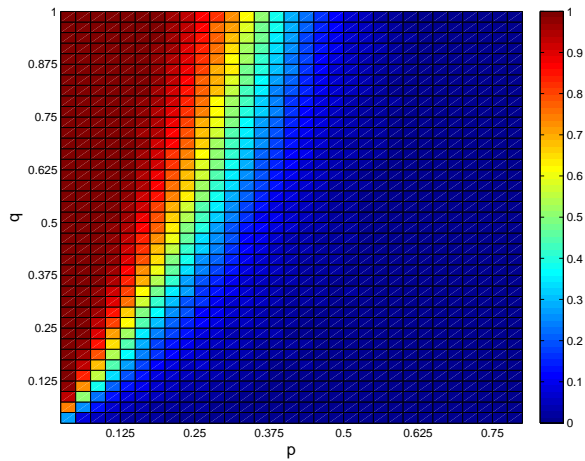}% Here is how to import EPS art
\caption{\label{fig:5} The cumulative probability for a finite epidemic range for \cite{collab}. Each point is obtained by averaging over 2000 simulations.} %range_integral_2d.eps}
\end{figure}

\begin{figure}[t]
\centering
\includegraphics*[width=0.5\textwidth]{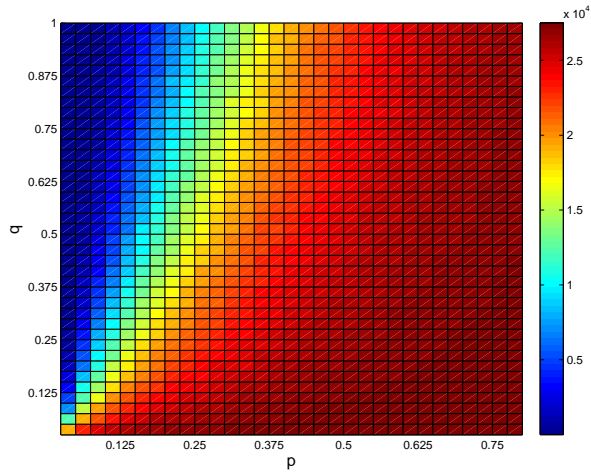}% Here is how to import EPS art
\caption{\label{fig:12} The average number of infected nodes for the network \cite{collab}. Each point is obtained by averaging over 2000 simulations.}
\end{figure}

\begin{figure}[t]
\centering
\includegraphics*[width=0.5\textwidth]{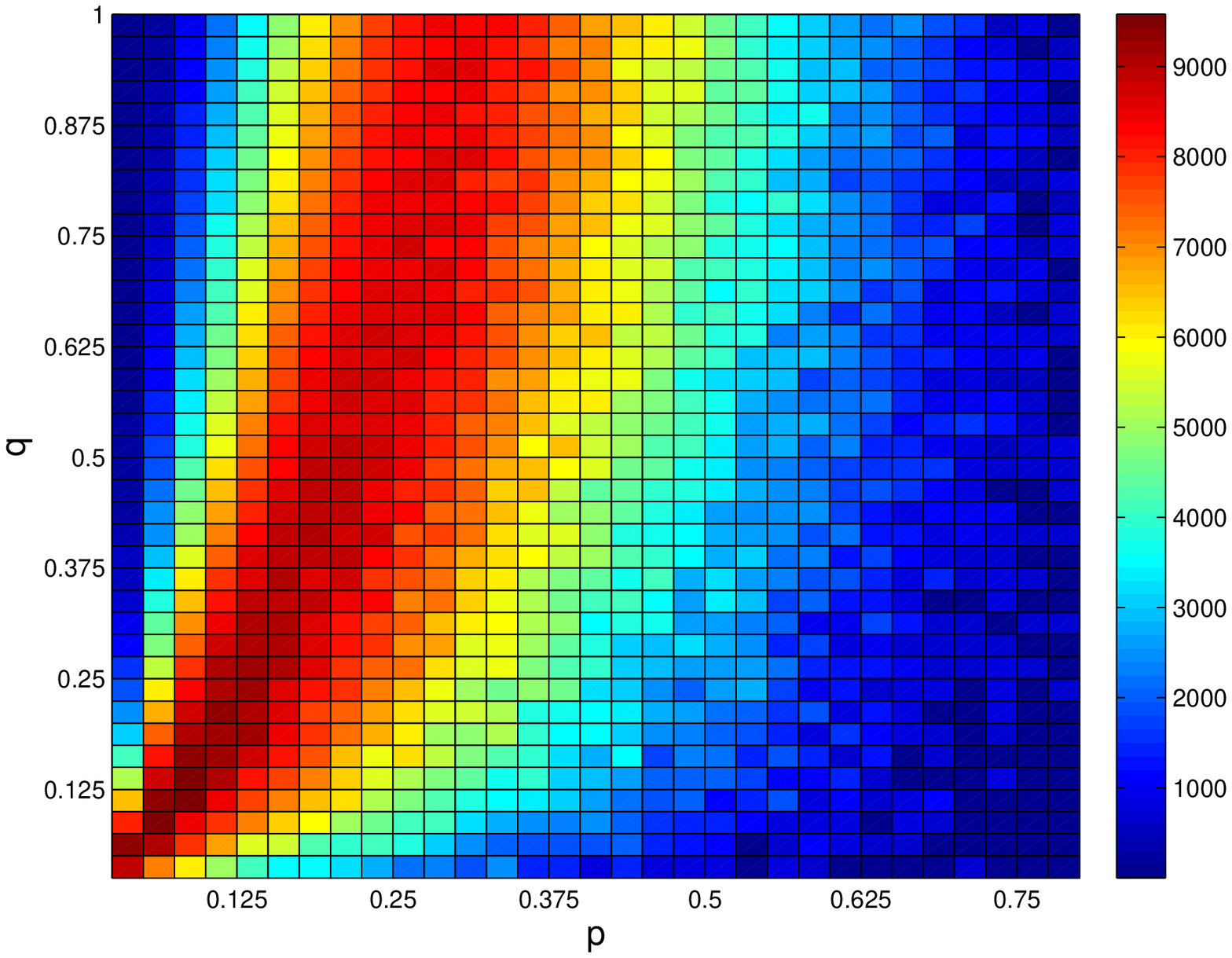}% Here is how to import EPS art
\caption{\label{fig:7} The standard deviation of the number of infected nodes for the network \cite{collab}. Each point is obtained by averaging over 2000 simulations.}
\end{figure}

\begin{figure}[t]
\centering
\includegraphics*[width=0.45\textwidth]{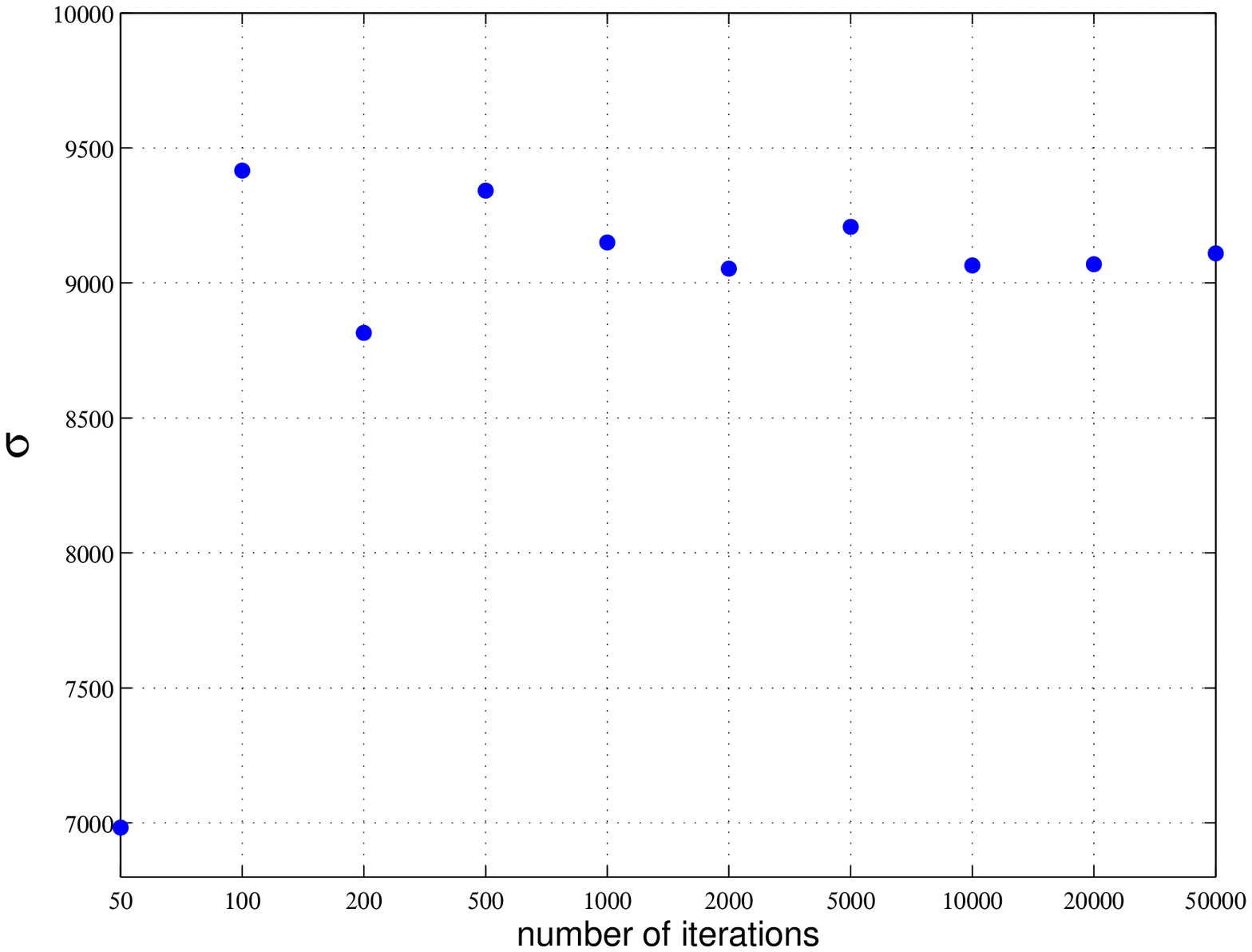}% Here is how to import EPS art
\caption{\label{fig:stdevstab} } The convergence of the standard deviation of the number of infected nodes for $p=0.125$ and $q=0.2$. The convergence of the standard deviation to a nonvanishing value is a strong indication of bimodal probability distribution.
\end{figure}

\begin{figure}[t]
\centering
\includegraphics*[width=0.5\textwidth]{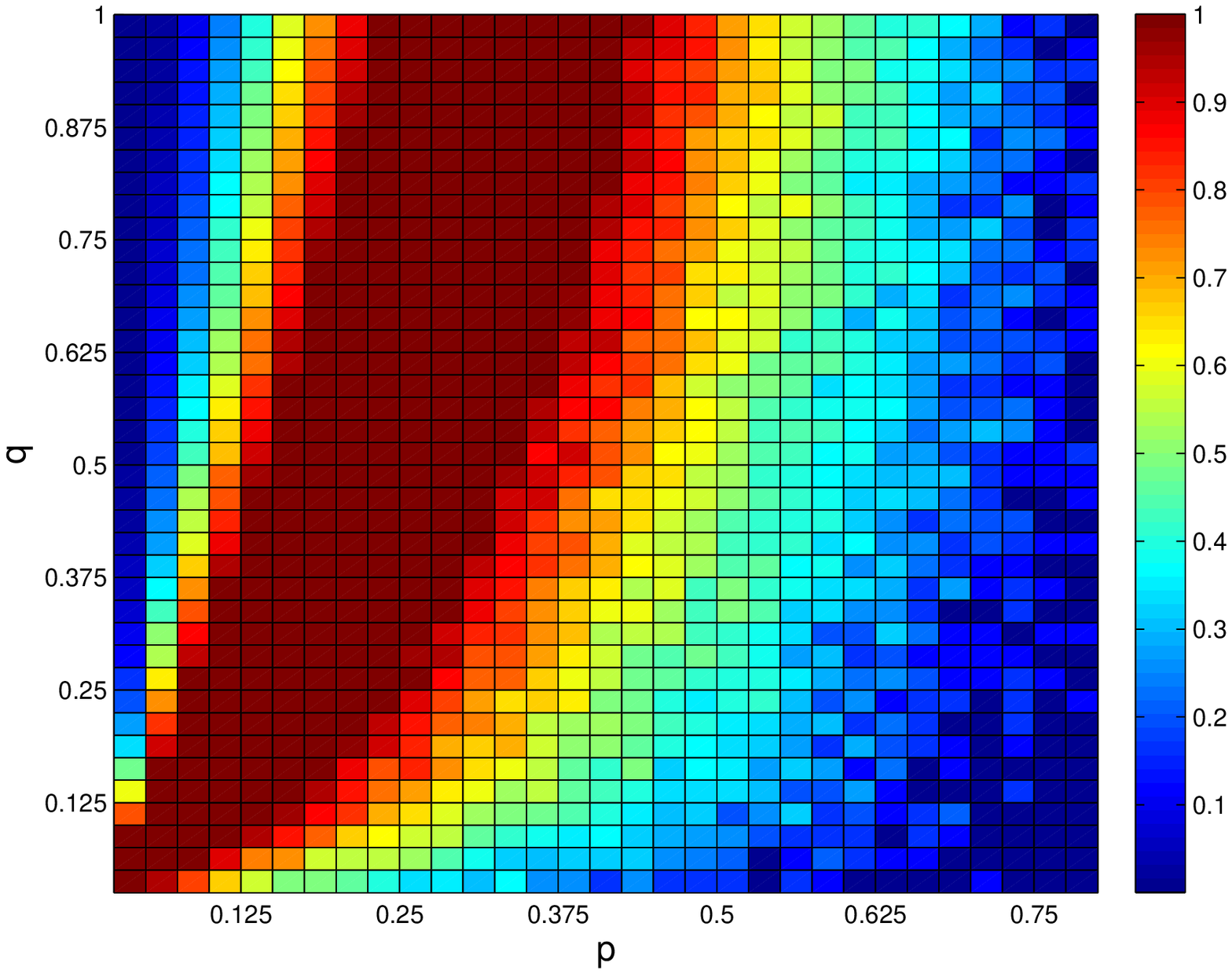}% Here is how to import EPS art
\caption{\label{fig:9} The length of a normalized $\pm3$ standard deviation interval of the number of infected nodes for the network \cite{collab}. Each point is obtained by averaging over 2000 simulations.}
\end{figure}

A crucial question is whether, for the chosen complex network and the initially infected node, we have to fine-tune parameters $p$ and $q$ to produce the bimodal probability distributions of the variables measuring the extent of disease spreading. To answer this question, we introduce a study of the entire $(p,q)$ parametric space of the SIR model: a $[0,1] \times [0,1]$ square. For each set of $(p,q)$ values we plot the value of the variable measuring the extent of disease spreading. For reasons which will soon be evident, we call the obtained plots {\em the phase diagrams of epidemic spreading}. An example of a phase diagram for the epidemic range is given in Fig. \ref{fig:5}. The quantity presented in the phase diagram is the cumulative probability for a finite epidemic range. For a complex network of finite size the practical method of calculation of the said cumulative probability is based on the bimodal character of the epidemic range distribution, see Fig. \ref{fig:3}. The cumulative probability is simply the sum of probabilities for the epidemic range starting from 0 up to the range where the probability drops to zero. From the phase diagram we can easily identify two extreme regimes. The first one is characterized by high $q$ and low $p$ where the cumulative probability for a finite epidemic range tends to 1. The second regime appears at high $p$ and low $q$ and there the cumulative probability for a finite epidemic range tends to 0. The existence of these two regimes in their respective ranges of $(p,q)$ parameters is not surprising having in mind the very meaning of these parameters. However, the phase diagram reveals a non-negligible area in the parametric space in which the cumulative probability for finite range differs significantly from both 0 and 1. This area is the transitional area connecting two extreme regimes of local containment and epidemic. Within this region it is comparably probable that the disease spreading would be contained or that it would explode into an epidemic. In Figs. \ref{fig:12}-\ref{fig:9} we have phase diagrams for the number of infected nodes. In Fig \ref{fig:12} we see that at low $p$ and high $q$ the number of infected nodes is low compared to the total number of nodes in the network. On the other hand, in the area of high $p$ and low $q$ the number of infected nodes is of the order of the total number of nodes in the network. The areas of the parametric space in which these two extreme regimes are realized are connected by a broad transitional area. A comparison of phase diagrams in Figs \ref{fig:5} and \ref{fig:12} shows that the transitional areas in both phase diagram correspond closely. The phase diagram for standard deviation of the number of infected nodes, depicted in Fig. \ref{fig:7} reveals that the standard deviation is large in the very transitional area which has been identified in Figs \ref{fig:5} and \ref{fig:12}. It should be stressed that this result is not a result of an insufficiently large sample of simulations. In the transitional area of the parametric space the standard deviation of the number of infected nodes saturates at a finite value as the size of sample increases, see Fig. \ref{fig:stdevstab}. This observation is fully consistent with a bimodal character of the probability distribution for the number of infected nodes. Finally, in Fig. \ref{fig:9} we present length of a normalized $\pm3$ standard deviation interval of the number of infected nodes defined as $(l,u)=(\mathrm{max} (0, (E(Y)-3 \sigma(Y))/N) ,\mathrm{min} ((E(Y) + 3 \sigma(Y))/N , 1))$, where $Y$ is the random variable of the number of infected nodes and $N$ is the total number of nodes in the network. According to ineqality of Chebyshev the probability of $Y$ taking value in the interval $(l,u)$ is at least $88.89\%$.

%{\bf this all indicates that we have different representations of the same bimodal dynamics!!!!}
%Such a plot provides the answer to the question how generic the bimodal behavior is. We call the obtained surfaces {\em the phase diagrams of disease spreading}. 
An important result, presented in phase diagrams given in Figs \ref{fig:5}-\ref{fig:9} is that the area of parametric space characterized by the bimodal distribution of the epidemic range and the number of infected nodes is large. Therefore, the appearance of the described bimodal distributions in phase diagrams is generic. 

Given the large degree of heterogeneity of empirical complex networks, the following important question is how these phase diagrams of epidemic spreading depend on the choice of an initially infected node. Preliminary simulations show that large differences may exist. Still, observed phase diagrams for initially infected nodes of very different degrees show important similarities and we could roughly characterize them as qualitatively the same. This important question also has a considerable practical importance since the choice of the initially infected node may describe the difference between a random outbreak (any randomly selected node) and a terrorist act (hub). A more elaborate study of the influence of the selection of the initially infected node is left for a future dedicated work \cite{mi2}.

%Description of diagnostic tools that we use - epidemic range and the number of infected nodes (both as random variables - we present their distributions)
%Realistic networks - which networks (we take small networks -several) - 
%which graphs (for the same network)
%1.range in unimod
%2.range in bimod
%3.numinfec in bimod 
%4.phase diagram for range (this could perhaps be commented in a footnote)
%5. phase diagram for range for several networks

%Demonstration of the unimodal/bimodal regimes. {\bf citation to Gallas} Introduction of the concept of phase diagram (how strict is this for realistic complex networks??).

%\begin{figure}[t]
%\centering
%\includegraphics*[width=0.5\textwidth]{std_num_inf_3d.eps}% Here is how to import EPS art
%\caption{\label{fig:8} The 3D plot of the standard deviation of the number of infected nodes for the network \cite{collab}. Each point is obtained by averaging over 1000 simulations.}
%\end{figure}

\begin{figure*}[t]
\centering
\includegraphics*[width=1.0\textwidth]{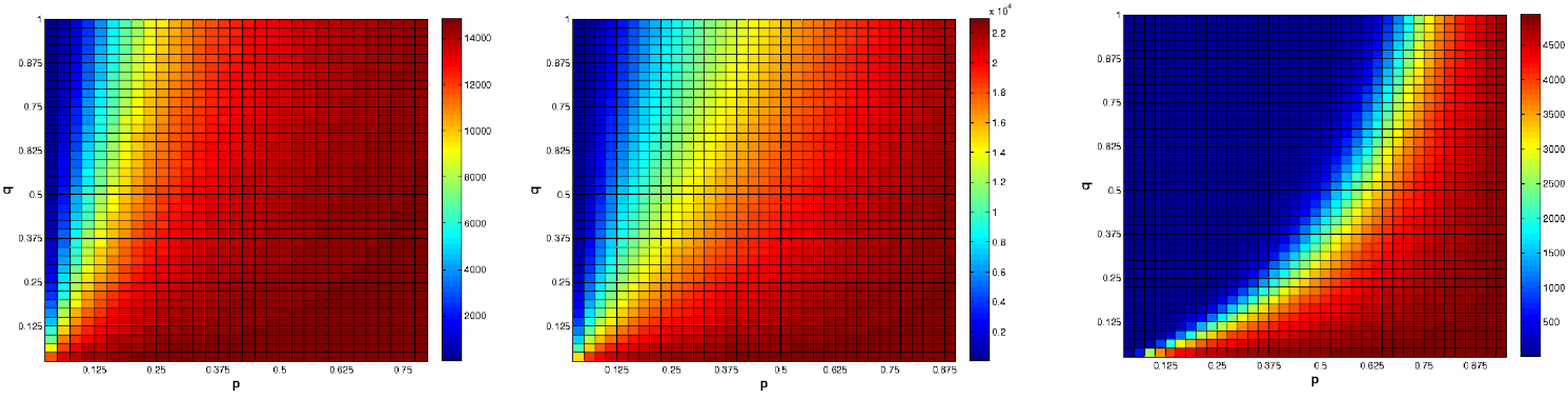}% Here is how to import EPS art
\caption{\label{fig:avg_num_inf_astro_internet_power_grid.eps} The phase diagram of epidemic spreading for the average number of infected nodes for co-authorships in astrophysics \cite{collab} (left), Internet at the level of autonomous systems \cite{net3} (center) and  the US Western States Power Grid \cite{net1} (right).}
\end{figure*}

Finally, there remains the question of the dependence of the presented results on the particular empirical network used in the discussion up to this point. The described analysis was repeated on several other empirical networks: An undirected, unweighted network representing the topology of the US Western States Power Grid \cite{net1}, network of coauthorships between scientists posting preprints on the Astrophysics E-Print Archive between Jan 1, 1995 and December 31, 1999.  \cite{collab}, and a a symmetrized snapshot of the structure of the Internet at the level of autonomous systems, reconstructed from BGP tables posted by the University of Oregon Route Views Project \cite{net3}. The form of phase diagrams observed in the study of \cite{collab} is the same as in simulations on other three empirical complex networks \cite{net1,collab,net3} though there are also notable differences in quantitative details, as depicted in Fig. \ref{fig:avg_num_inf_astro_internet_power_grid.eps}. The conclusion is that the qualitative features of the phase diagrams of disease spreading, are the same for the SIR disease spreading model on the different studied empirical complex networks. This finding is a strong indication that the form of phase diagrams is generic across complex networks and that it reflects some fundamental dynamics of disease spreading.

\section{Theoretical model for m-ary trees} 

The observed pattern of epidemic spreading calls for identification of the underlying mechanism producing it.  The structure of empirical complex networks, despite many generic properties, is very intricate. Many structural characteristics of complex networks might contribute to the pattern of epidemic spreading. Therefore it is reasonable to start from a simple structure that still incorporates the key network properties for the process of disease spreading. 
Many complex networks can be locally well approximated by a tree-like structure. In empirical networks we very rarely observe tree-like structures and we say that the local description of the network as a tree is a good approximation if the clustering is small. 
%({\bf ubaciti reference - Vinko}). 
The IT/communication empirical networks in general fit into this class, such as e-mail and Internet router networks \cite{it}. The social contact networks tend to be more clustered, although exceptions exist, such as coauthorship networks in medicine  \cite{med}. In this paper we exploit these findings and consider the disease spreading on a m-ary tree as a starting model in the study of the phase diagram of epidemic spreading. In m-ary trees that we consider in this paper each node has $m$ children. The simplicity of this starting model provides a transparent analysis and also probes the relevance of the very tree structure as the backbone of the complex network structure in the disease spreading dynamics \cite{treestruct}. 
%Furthermore, there are several reasons why study of epidemic spreading on regular trees is more practically relevant. The first typical regime of the epidemic ``die-out'' is usually realized within a small distance from the initially infected node. This regime, therefore, is fully realized within the portion of the network which is well described by the tree. 
An additional argument for considering the disease spreading on tree-like networks is that the results obtained in this setting may serve as a useful lower bounds for the extent of the disease spreading (expressed i.e. as a number of infected nodes or the range of the epidemic). In general, the number of infected nodes and the range of the epidemic on a given empirical network is always larger than for an epidemic on any spanning tree with the same initial node. The extent of the disease spreading on trees, expressed i.e. as a number of infected nodes or the range of the epidemic,  represent a lower bound for the extent of the disease spreading on a complex network. This fact becomes especially useful when the result for a tree reaches its maximally allowed value. For example, if the probability for spreading to entire tree is 1, the corresponding probability for a complex network will also be 1. 
%({\bf modificirati????}). {\bf Jos koji argument???) Ubaciti sliku binarnog stabla.}.    

%Regular tree model. Why regular tree? (Alen's argument??)
%The figure of a binary tree.

\begin{figure}[t]
\centering
\includegraphics*[width=0.5\textwidth]{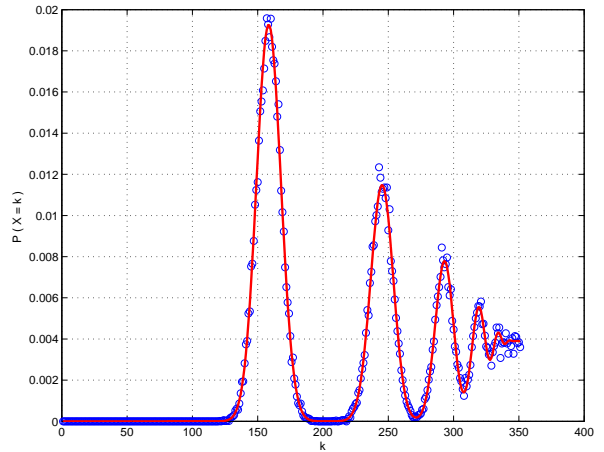}% Here is how to import EPS art
\caption{\label{fig:distr} The probability distribution $p_{n,k}^{(s)}$ for $p=0.45$, $q=0.45$, $s=1$ and $n=350$ averaged over $50 000$ simulations. The results of simulations are depicted by circles and the full line represents the analytically obtained function (\ref{eq:distrX}). }
\end{figure}

The study of the epidemic spreading on a m-ary tree is performed both analytically and using computer simulations. The basic element for describing the epidemic on a m-ary tree analytically is the probability distribution of the infected node infecting some number of its children nodes. Let us consider a problem of epidemic spreading in a complete bipartite graph consisting of two classes of nodes: $s$ infected nodes (class I) and $n$ susceptible nodes (class II). Each node from the class I is connected to all nodes from the class II. We are interested in a number of susceptible nodes that will get infected during the course of the epidemic. The random variable of the number of nodes in class II that eventually get infected is denoted by $X_n^{(s)}$. The probability of $k \le n$ nodes of class II getting eventually infected is given by the expression
\begin{eqnarray}
\label{eq:distrX}
&p^{(s)}_{n,k}& \equiv P(X_n^{(s)}=k) = \nonumber \\&=& q^s \binom{n}{k} \sum_{l=0}^k \binom{k}{l} (-1)^l \left(\frac{(1-p)^{n-k+l}}{1-(1-q)(1-p)^{n-k+l}}\right)^s \, .
\end{eqnarray}
This distribution exhibits very interesting properties, such as multiple peaks as depicted in Fig. \ref{fig:distr}.

%$m$ to $n$ infection probability distribution. 

The probability distribution for the range of the epidemic spreading on a m-ary tree can be calculated within the theory of branching processes \cite{Feller}. The random variable $Z_n$ representing the number of infected nodes at the $n^{th}$ level of the tree  ($n > 1$) can be represented as 
\begin{equation}
\label{eq:Zn}
Z_n= \sum_{i=1}^{Z_1} Z_{n,i} \, ,
\end{equation}
where $Z_{n,i}$ is the random variable of a number of $n^{th}$ level infected nodes that got infected via the $i^{th}$ infected node at the first level. At the first level we have $Z_1 \sim X_{m}^{(1)}$. The generating functions $F_n(s)=\sum_0^{\infty} P(Z_n=l) s^l$ are found to satisfy the relation
\begin{equation}
\label{eq:rec} F_n(s)=F(F_{n-1}(s)) \, , 
\end{equation}
where $F(s)=F_1(s)$. 
The probability that there are no infected nodes (at the end of the epidemic) at the $n^{th}$ level, $F_n(0)$, is the probability that the range of the epidemic is $\le n-1$. Then the range of the epidemic is $n$ with a probability
\begin{equation}
\label{eq:probrange} d_n=F_{n+1}(0)-F_n(0) \, .
\end{equation}
For $n \ge 1$ the quantities $P_n(0)$ can be calculated iteratively from (\ref{eq:rec}) with $d_0=F(0)=p_{m,0}^{(1)}$. An excellent agreement of analytic and simulational results for a typical distribution of the range is found. 

\begin{figure}[t]
\centering
\includegraphics*[width=0.5\textwidth]{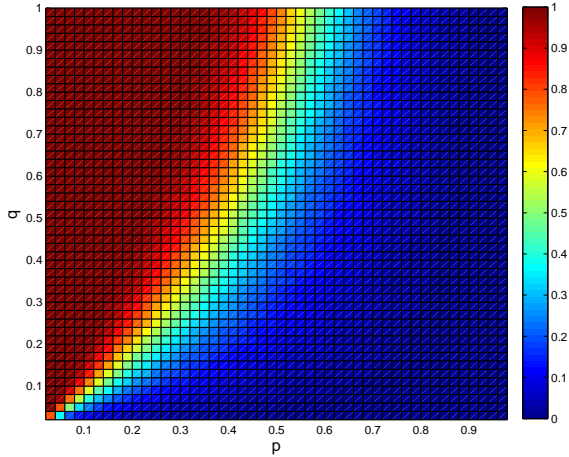}% Here is how to import EPS art
\caption{\label{fig:4} The cumulative probability for a finite epidemic range for a binary tree with 12 levels. Each point is obtained by averaging over 10000 simulations.}
\end{figure}

\begin{figure*}[t]
\centering
\includegraphics*[width=1.0\textwidth]{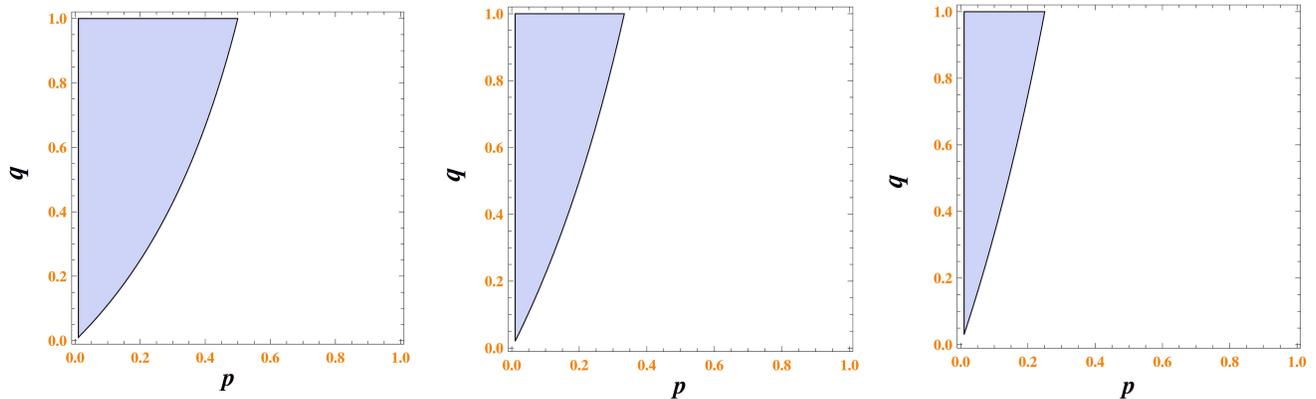}% Here is how to import EPS art
\caption{\label{fig:varm} Phase diagrams of epidemic spreading for m-ary trees for $m=2$ (left), $m=3$ (center), and $m=4$ (right) obtained from analytical considerations. The shaded region corresponds to a unimodal behavior and the unshaded region to the bimodal behavior.  }
\end{figure*}

The cumulative probability of having a finite range of epidemic $d_{tot} \equiv \sum_{i<\infty} d_i$ can be used for the definition of the phase diagram of epidemic spreading on a m-ary tree. The nature of the solutions of the equation $F(x_f)=x_f$ serves as an equivalent tool for the definition of phases in the said phase diagram. For $d_{tot}=1$, for which a solution $x_f<1$ does not exist, the range distribution is unimodal and the disease is locally contained. For $d_{tot}<1$, where the solution $x_f<1$ exists, the range distribution is bimodal and there are finite probabilities for the locally contained outbreak and for the epidemic sweeping through the entire tree. A sharp boundary between these two phases exists and it can be obtained from the condition $E(X_m^{(1)})=m E(X_1^{(1)})=1$. An interesting consequence of this relation for $m=1$ is that for all values $p<1$ there exits only the phase of local containment of the disease. The phase diagram of the epidemic spreading on a m-ary tree obtained in simulations is depicted in Fig. \ref{fig:4} whereas those obtained from analytical considerations for m-ary trees are given in Fig. \ref{fig:varm}. 

\begin{figure}[t]
\centering
\includegraphics*[width=0.5\textwidth]{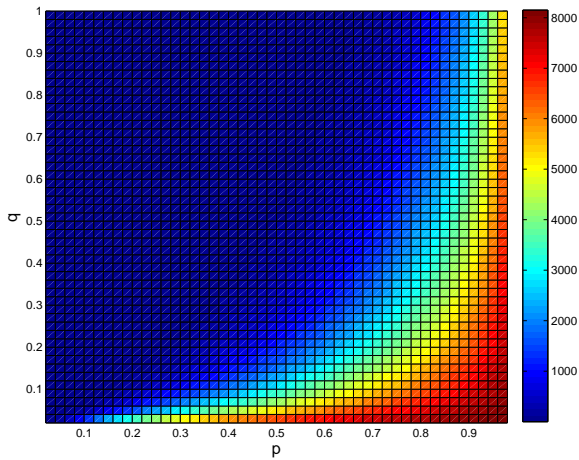}% Here is how to import EPS art
\caption{\label{fig:11} The average number of infected nodes for a binary tree with 12 levels. Each point is obtained by averaging over 10000 simulations.}
\end{figure}

\begin{figure}[t]
\centering
\includegraphics*[width=0.5\textwidth]{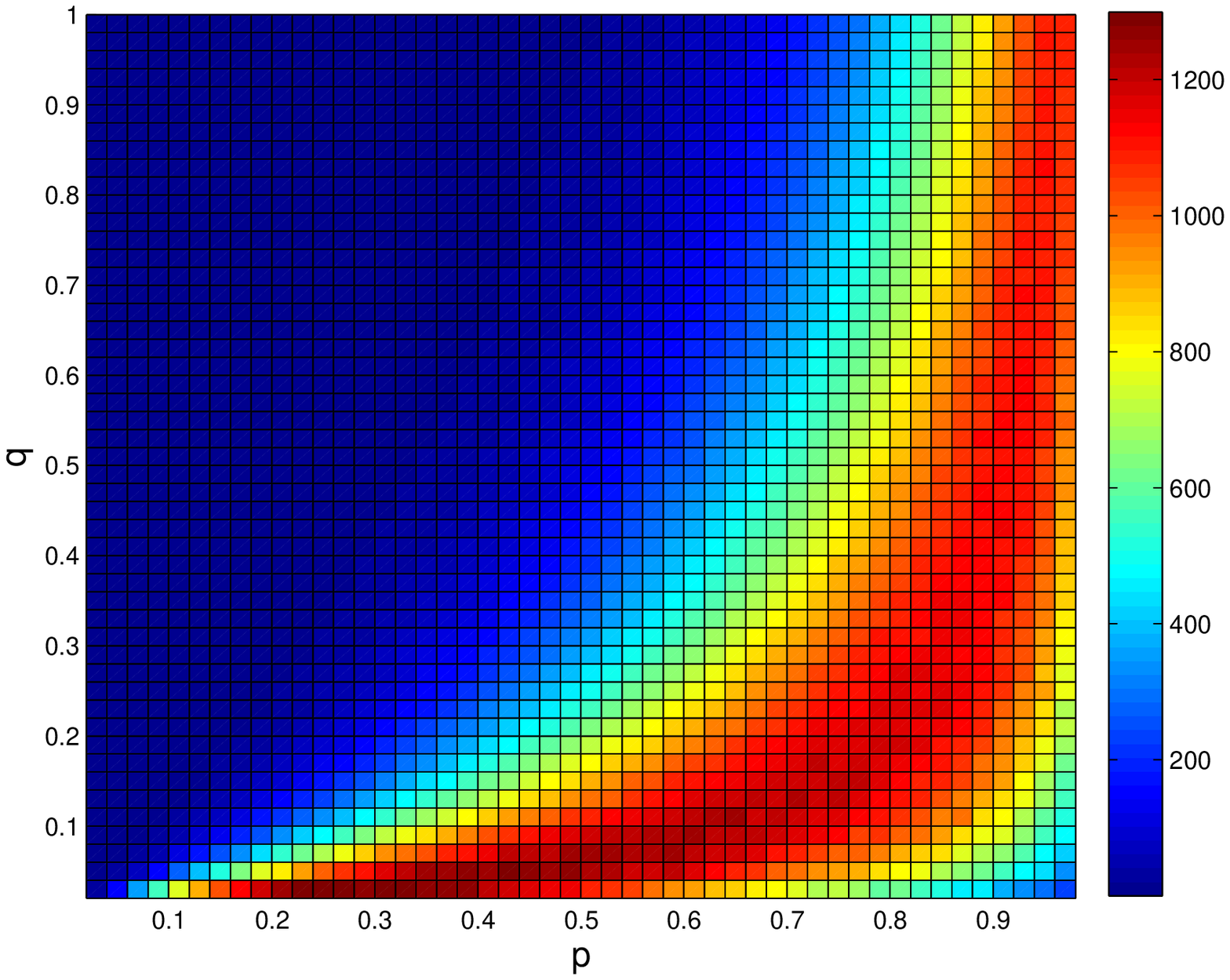}% Here is how to import EPS art
\caption{\label{fig:6} The standard deviation of the number of infected nodes for a binary tree with 12 levels. Each point is obtained by averaging over 10000 simulations.}
\end{figure}

\begin{figure}[t]
\centering
\includegraphics*[width=0.5\textwidth]{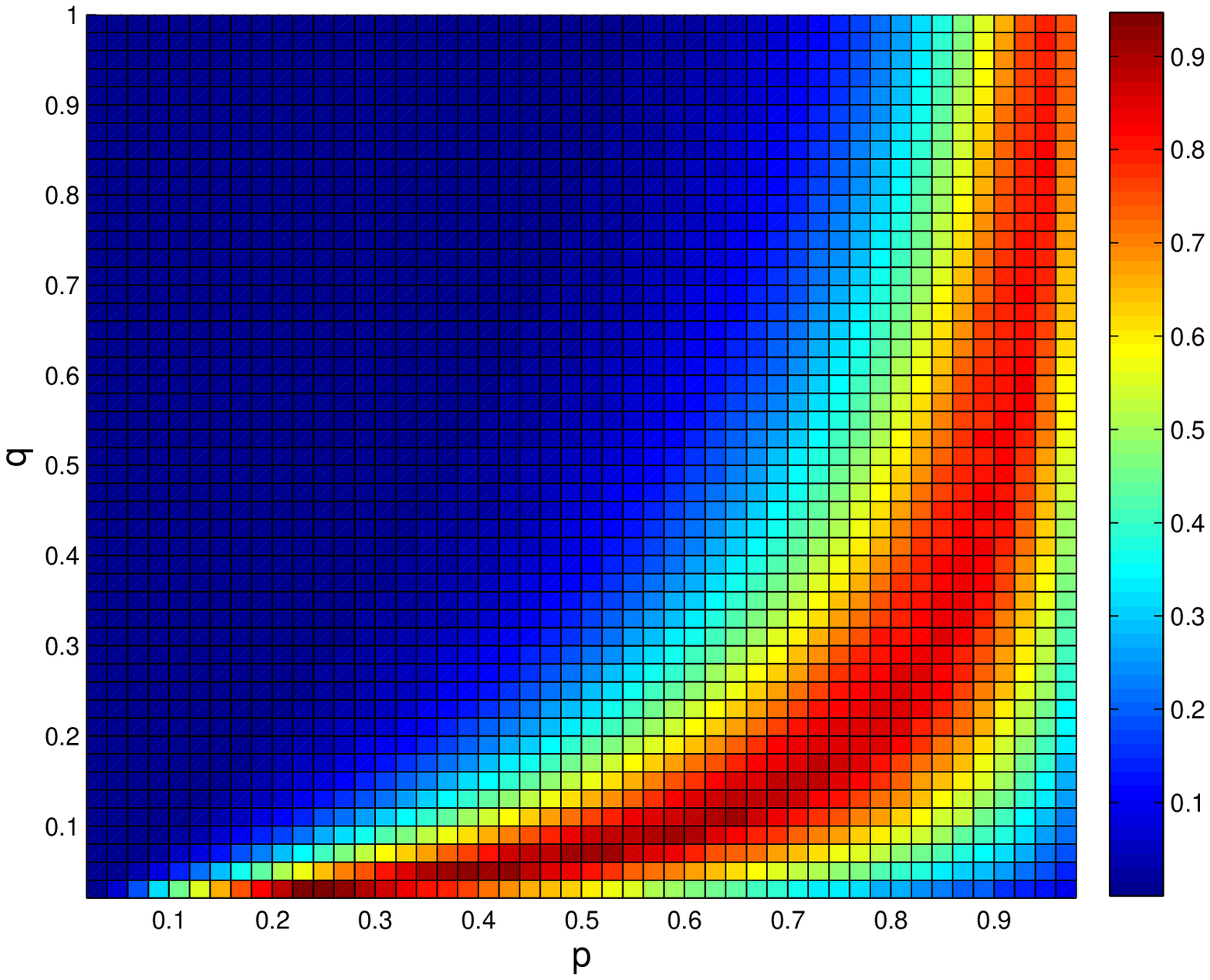}% Here is how to import EPS art
\caption{\label{fig:10} The normalized $\pm3$ standard deviation interval for the number of infected nodes for a binary tree with 12 levels. Each point is obtained by averaging over 10000 simulations.}
\end{figure}

%The branching proces. Generating functions. Derivation of the range distribution.
%\cite{Feller}. Description of unimodal vs. bimodal. Line of separation. Comparison with realistic networks.
%The phase diagrams for various values of $m$. ($m=2,3,6$??)   
%Simulations vs. analytic description (plots).

The total number of infected nodes in a m-ary tree with $n$ levels can be represented by a variable $Y_n=\sum_{i=0}^{n} Z_i$
The analytic expressions for the expectation and the variance of the number of infected nodes $Y_n$ are given by the following expressions
\begin{equation}
\label{eq:treeavgn}
E(Y_n)= \frac{\left(m p_{1,1}^{(1)}\right)^{n+1}-1}{m p_{1,1}^{(1)}-1} \, 
\end{equation}
and
\begin{eqnarray}
\label{eq:treestdn} 
 Var(Y_n) &=& \frac{m \left( (m-1) p_{2,2}^{(1)}-\left( m p_{1,1}^{(1)}-1 \right) p_{1,1}^{(1)} \right)}{\left( m p_{1,1}^{(1)}-1 \right)^3 } \nonumber  \\
&\times& 
\left[ \left( m p_{1,1}^{(1)} \right)^{2 n+1}-(2 n+1) \left( m p_{1,1}^{(1)} \right)^{n+1} \right. \nonumber \\
&& \left. + (2 n+1) \left( m p_{1,1}^{(1)} \right)^{n}-1 \right] \, . 
\end{eqnarray}

These analytical results are in excellent agreement with the simulations on  m-ary trees. The phase diagrams for the average number, standard deviation and the normalized $\pm 3\sigma$ interval of infected nodes are presented in Figures \ref{fig:11}, \ref{fig:6}, and \ref{fig:10} respectively. 
 
%The number of infected. The formulae for the average and stdev. 
%Probability distribution (for a tree??????) Which plots??

\section{Discussion and conclusions}

The main result of the analysis of the disease spreading on regular trees is that they reproduce the main qualitative characteristics of the phase diagrams of disease spreading observed on empirical complex networks. The regions of high $q$ and low $p$ where the local containment of the disease dominates and the region of low $q$ and high $p$ where the onset of epidemic is almost certain are connected by a large transitional area where the local containment and the epidemic spread have comparable probabilities. The insight provided by the analysis of regular trees suggests the following mechanism producing the observed phase diagrams of disease spreading. The process of the disease spreading is an interplay of two driving forces: the stochastic nature of the SIR model (and other epidemiological models) which always allows the possibility the disease will not be propagated further at any step (as an extreme demonstration, one should note that for a m-ary tree with $m=1$ the spreading of the disease is always contained) and the exponentially growing number of paths through which the disease may spread. When the extinguishing nature of the SIR model dominates the local containment prevails. A very large number of disease spreading pathways on the other hand strongly stimulates the onset of epidemic. A situation in which these two driving forces are largely in equilibrium is realized in the transitional area where both typical regimes are comparably probable. 
%{\bf plots with m=1, m=2, m=3...???} 

The disease spreading at a m-ary tree in SIR model exhibits the same qualitative or even semi-quantitative features of phase diagrams as those observed in empirical complex networks. This is an important result since it indicates that the structure of the m-ary tree is sufficient to reproduce the main features of the phase diagrams of epidemic spreading. At a quantitative level, however, there are many peculiarities that are observed in the disease spreading on empirical networks. A reasonable path in research efforts to explain these peculiarities is studying the effects or network structural features more complex that the underlying (spanning) trees. The study of the effects of nontrivial degree distributions and the influence of cycles are natural first stops along this path. 

\section{Acknowledgements}

The work of M. \v{S}. is financed by Ministry of Education Science and Sports of the Republic of Croatia under Contract No. 036-0362214-1987 and 098-1191344-2860. The work of H. \v{S}. is supported by the Ministry of Education Science and Sports of the Republic of Croatia under Contract No. 098-0352828-2863.

%Discussion and Conclusions

%1. Is the description in terms of a regular tree acceptable qualitatively and quantitatively

%2. What happens for a stochastic tree (that is the case of a $m=1$ BA model)

%3. The role of cycles

%\begin{figure}
%\includegraphics{fig_1}% Here is how to import EPS art
%\caption{\label{fig:epsart} A figure caption. The figure captions are
%automatically numbered.}
%\end{figure}

%\begin{figure}[t]
%\centering
%\includegraphics*[width=0.5\textwidth]{KiKrVsp.eps}
%\caption{\label{KiKrVsp.eps} The change of expected initial correlations of onevertex degrees $\langle k_ik_r \rangle$ calculated from the inverse of transformation matrix $\mathbf{T}^{-1}_{1v}$ for three different Wikipedias. Expected values of monitored correlations are changing the sign for the value of parameter about $p\sim 0.07$. Since the product moment correlations are strictly positive, this behavior indicates that there is just a small fraction of reciprocal arcs which are degree independent. In the case of Wikipedias the maximal value of parameter $p$ is about $p\sim 0.16$. }
%\end{figure}

%Fig.~\ref{fig:wide} is a figure that is too wide for a single column,
%so instead the \texttt{figure*} environment has been used.

%\begin{figure*}
%\includegraphics{fig_2}% Here is how to import EPS art
%\caption{\label{fig:wide}Use the figure* environment to get a wide
%figure that spans the page in \texttt{twocolumn} formatting.}
%\end{figure*}

%\newpage %Just because of unusual number of tables stacked at end

\bibliography{epitree}% Produces the bibliography via BibTeX.

\end{document}